\documentclass[12pt,preprint]{aastex}
\begin{document}

\title{High Energy Emission from the Prompt Gamma-Ray Burst}

\author{Dafne Guetta\altaffilmark{1} and Jonathan Granot\altaffilmark{2}}

\altaffiltext{1}{Osservatorio astrofisico di Arcetri, L.E. Fermi 2, Firenze, Italy; dafne@arcetri.astro.it}
\altaffiltext{2}{Institute for Advanced Study, Olden Lane, Princeton, NJ 08540; granot@ias.edu}

\begin{abstract}

We study the synchrotron and synchrotron self-Compton (SSC) emission from
internal shocks that are responsible for the prompt $\gamma$-ray emission
in Gamma-Ray Bursts (GRBs), and consider the relation between these two
components, taking into account the high energy cutoff due to pair
production and Thomson scattering. We find that in order for the peak energy
of the synchrotron to be $E_p\sim 300\;{\rm keV}$ with a variability time
$t_v\gtrsim 1\;{\rm ms}$, a Lorentz factor of $\Gamma\lesssim 350$ is needed,
implying no high energy emission above $\sim 30\;{\rm MeV}$ and the synchrotron
component would dominate at all energies. If we want both
$E_p\sim 300\;{\rm keV}$ and prompt high energy emission up to $\sim 2\;{\rm GeV}$,
as detected by EGRET for GRB 940217, we need $\Gamma\sim 600$ and
$t_v\sim 0.1\;{\rm ms}$, which might be resolved by super AGILE. If such prompt
high energy emission is common in GRBs, as may be tested by GLAST, then for
$t_v\gtrsim 1\;{\rm ms}$ we need $\Gamma\gtrsim 350$, which implies
$E_p\lesssim 100\;{\rm keV}$. Therefore if X-ray flashes are GRBs with high
values of $t_v$ and $\Gamma$, they should produce $\gtrsim 1\;$GeV emission.
For an electron power law index $p>2$, the SSC component dominates the emission above
$\sim 100\;{\rm MeV}$. Future observations by GLAST may help determine the value of
$p$ and whether the high energy emission is consistent with a single power law
(implying one component--the synchrotron, is dominant) or has a break where the
$\nu F_\nu$ slope turns from negative to positive, which implies that the SSC
component becomes dominant above $\sim 100\;$MeV. The high energy emission is expected
to show  similar variability and time structure to that of the soft $\gamma$-ray
emission. Finally, we find that in order to see delayed high energy emission from the
prompt GRB due to pair production with the cosmic IR background, extremely small
inter-galactic magnetic fields ($\lesssim 10^{-22}\;{\rm G}$) are required.

\end{abstract}

\keywords{gamma rays: bursts---ISM: jets and outflows---
radiation mechanisms: nonthermal}

%%%%%%%%%%%%%%%%%%%%%%%%%%%%%%%%%%%%%%%%%%%%%%%%%%%%%%%%%%%%%%%%
\section{Introduction}
\label{sec:intro}

The leading models of Gamma-Ray Bursts (GRBs) involve a relativistic flow emanating
from a compact central source, where the prompt gamma-ray emission is attributed to
internal shocks within the outflow itself, that arise from variability in its Lorentz
factor, while the afterglow results from an external shock that is driven into
the ambient medium, as it decelerates the original ejecta (Rees \& M\'esz\'aros 1994;
Sari \& Piran 1997). In this so called `internal-external'
shock model, the duration of the prompt GRB is directly related to the time during which
the central source is active. The most popular emission mechanism is synchrotron radiation
from relativistic electrons accelerated in the shocks, that radiate in the strong magnetic fields
(close to equipartition values) within the shocked plasma. An additional radiation mechanism that
may also play an important role is synchrotron self-Compton (SSC), which is the upscattering
of the synchrotron photons by the relativistic electrons, to much higher energies.

The synchrotron and SSC components from internal shocks have been
considered in previous works in various contexts. Papathanassiou
\& M\,esz\,aros (1996) studied the emission from internal shocks,
focusing on the comparison between internal and external shocks.
Pilla \& Loeb (1998) calculated the spectrum from internal shocks
taking into account multiple Compton scattering and pair
production. They show the broad band spectrum for a fixed radius
of collision, $R$, and varying Lorentz factor, $\Gamma$, of the
outflow, and for a fixed $\Gamma$ and a varying $R$. Our treatment
differs in that we assume different free parameters, namely the
Lorentz factor $\Gamma$ and the variability time, $t_v$ of the
central engine that emits the outflow, rather than the $\Gamma$
and $R$. Under our assumptions, the radius of collision scales as
$R\sim 2\Gamma^2ct_v\propto\Gamma^2$, and is not independent of
$\Gamma$. This results in different conclusions as to the relation
between the prompt gamma-ray emission in the BATSE range and the
emission at higher energies, which is the main subject of our
work. Panaitescu \& M\,esz\,aros (2000) explored the possibility
that the prompt gamma-ray emission in the BATSE range arises from
the SSC component rather than the synchrotron, where the latter is
in the optical or UV range. Recently, Dai \& Lu studied the SSC
emission from internal shocks, concentrating on the possible
interaction of high energy photons with the IR background (see
section \S \ref{limits}).

In this {\it Letter} we calculate the high energy emission during
the prompt GRB from internal shocks, for both the synchrotron and
SSC components, and consider the relations between these two
components. We estimate the high energy cutoff and study the
constraints on the model parameters that arise from the
requirement that $E_p$ (the typical photon energy of the
synchrotron component) will be in the BATSE range.

The synchrotron and SSC spectra are calculated in \S \ref{IS}, and expressions are provided for
the break frequencies and flux normalization. In \S \ref{limits} we derive the constraints on the model
that arise from the optical depth to pair production and to Thomson scattering. We consider the
recent claim for a possible delayed emission due to the pair production of high energy photons
with the IR background (Dai \& Lu 2002) and we show that in order for this radiation to be
detectable, a very small ($\lesssim 10^{-22}\;{\rm G}$) inter-galactic magnetic field is needed.
Our results are discussed in \S \ref{sec:dis}.

%%%%%%%%%%%%%%%%%%%%%%%%%%%%%%%%%%%%%%%%%%%%%%%%%%%%%%%%%%%%%%%%
\section{The Synchrotron and SSC Emission}
\label{IS}

The Lorentz factor of the flow, $\Gamma$, is assumed to vary on a time scale $t_v$ and with
a typical amplitude of $\Delta\Gamma\sim\Gamma$. The collisions between the shells
typically occur at a radius
\begin{equation}\label{R}
R\approx 2\Gamma^2ct_v=6\times 10^{13}\,\Gamma_{2.5}^2 t_{v,-2}\;{\rm cm}\ ,
\end{equation}
where $\Gamma_{2.5}=\Gamma/10^{2.5}$ and
$t_{v,-2}=t_v/10^{-2}\;{\rm s}$. The internal energy that is
released in each collision between shells is distributed among
electrons magnetic fields and protons, with fractions
$\epsilon_e$, $\epsilon_B$ and $1-\epsilon_e-\epsilon_B$,
respectively. Since the electrons are typically in the fast
cooling regime (as is shown below), the luminosity
$L=(\Omega_j/4\pi)L_{iso}$ (where $\Omega_j$ is the solid of the
GRB outflow) of a single pulse, that corresponds to a single
collision between two shells within the outflow, is equal to the
rate at which energy is given to the electrons by the internal
shock.\footnote{The fact that we refer to the flux of a single
pulse, rather than the time average flux of all the pulses
throughout the burst, results in a factor of 3 difference in the
expression for the internal energy. However, as this kind of
calculation is anyway accurate only up to factors of order unity,
and similar uncertainties result from the unknown details of the
outflow, and factors of (at least) order unity are expected
between different collisions between shells in the same GRB,
shocked the exact choice of the parameterization of the luminosity
is not very important.} The internal shocks are expected to be at
least mildly relativistic, so that we may assume that in the rest
frame of the shocked fluid, the velocity of the shock is close to
that for a relativistic shock, $c/3$. The local frame luminosity
is $L'=L/\Gamma^2=\Omega_jR^2\epsilon_e e' c/3$, and the internal
energy density is given by
\begin{equation}\label{e}
e'\approx{3L_{iso}\over 4\pi R^2c\Gamma^2\epsilon_e}=2.2\times 10^8
\epsilon_{e}^{-1}L_{52}\Gamma_{2.5}^{-6}t_{v,-2}^{-2}\;{\rm erg\;cm^{-3}}\ ,
\end{equation}
where $L_{iso}=L_{52}10^{52}\,{\rm erg/s}$. Primed quantities are
measured in the local rest frame of the shocked fluid, while
un-primed quantities are measured in the observer frame.

For a mildly relativistic shock, the internal energy behind the
shock is similar to the rest energy,
\begin{equation}\label{n}
n'\approx{e'\over m_p c^2}\approx
{3L_{iso}\over 4\pi R^2m_pc^3\Gamma^2\epsilon_e}
=1.5\times 10^{11}\epsilon_{e}^{-1}L_{52}\Gamma_{2.5}^{-6}
t_{v,-2}^{-2}\;{\rm cm^{-3}}\ ,
\end{equation}
The magnetic field is
\begin{equation}\label{B}
B'=\sqrt{8\pi\epsilon_B e'}\approx 7.5\times 10^4\epsilon_{e}^{-1/2}\epsilon_B^{1/2}
L_{52}^{1/2}\Gamma_{2.5}^{-3}t_{v,-2}^{-1}\;{\rm G}\ .
\end{equation}
The electrons are accelerated in the shock to a power-law distribution of energies
$N(\gamma)=dn'/d\gamma \propto \gamma^{-p}$ for $\gamma$ in the range
$\gamma_m<\gamma<\gamma_M$. The minimum Lorentz factor is given by
\begin{equation}\label{gamma_m}
\gamma_{m}=\left({p-2\over p-1}\right){\epsilon_{e}e'\over n' m_e c^2}\approx
\left({p-2\over p-1}\right){m_p\over m_e}\,\epsilon_{e}\approx 610 f_p\,\epsilon_e\ .
\end{equation}
where $f_p\equiv 3(p-2)/(p-1)$.

The electrons radiatively cool by the combination of the synchrotron
and SSC processes, the timescales of which are
$t'_{\rm syn}\sim 6\pi m_e c/\sigma_T B'^2 \gamma_e$  and
$t_{SC}=t_{\rm syn}/Y$,the combined cooling time being
$t'_c=(1/t'_{\rm syn}+1/t'_{SC})^{-1}=t'_{\rm syn}/(1+Y)$, where
\begin{equation}\label{Y}
Y\approx\left\{\matrix{\epsilon_{e}/\epsilon_{B} &
\epsilon_{e}\ll\epsilon_{B} \cr & \cr
\sqrt{\epsilon_{e}/\epsilon_{B}} & \epsilon_{e}\gg\epsilon_{B}}\right. \ ,
\end{equation}
is the Compton y-parameter (Sari, Piran \& Narayan 1996).
The maximum Lorentz factor of the electrons, $\gamma_M$, and the cooling Lorentz factor,
$\gamma_c$, are set by equating $t'_c$ with the acceleration time, $\sim 2\pi \gamma_e m_e c/q B'$
(where $q$ is the electric charge of the electron), and the dynamical time,
$t'_{dyn}=R/c\Gamma\approx 2\Gamma t_v$, respectively:
\begin{eqnarray}
\gamma_{M}&=&\sqrt{3 q\over B'\sigma_T(1+Y)}\approx
1.7\times 10^5\,{\epsilon_{e}^{1/4}\Gamma_{2.5}^{3/2}t_{v,-2}^{1/2}\over
\sqrt{1+Y}\,\epsilon_B^{1/4} L_{52}^{1/4}}
\ ,\label{gamma_M}
\\
\gamma_{c}&=&{6\pi \Gamma m_e c^2\over (1+Y)B'^2 \sigma_T R}\approx
0.02\,{\epsilon_{e}\Gamma_{2.5}^{5}t_{v,-2}\over(1+Y)\epsilon_B L_{52}}\ .
\label{gamma_c}
\end{eqnarray}
Obviously, $\gamma_c<1$ no longer corresponds to the Lorentz factor of the electrons,
and instead represents the fraction of the dynamical time (the shell shock crossing
time for internal shocks) during which the electrons cool to a non-relativistic
random Lorentz factor. In this case, the electrons are relativistic only within
a thin layer behind the shock, of width $\gamma_c\Delta'$, where $\Delta'$ is
the width of the shell, and are cold (i.e. non-relativistic)
in the rest of the shell. As can be seen from equations (\ref{gamma_m}) and (\ref{gamma_c}),
typically $\gamma_c\ll\gamma_m$ and the electrons are fast cooling.

The synchrotron frequency and the total synchrotron power of a single electron are given by
\begin{equation}\label{syn}
\nu_{syn}=\nu_0\gamma_e^2=\Gamma\,{3q_eB'\gamma_e^2\over 16m_e c}\quad,\quad
P_{e,syn}=\Gamma^2{4\over3}\sigma_T c{B'^2\over 8\pi}\gamma_e^2\ ,
\end{equation}
where $\nu'_{syn}=\nu_{syn}/\Gamma$ and $P'_{e,syn}=P_{e,syn}/\Gamma^2$.
The self absorption frequency is typically ${\rm max}(\nu_c,\nu_0)<\nu_{sa}<\nu_m$,
and the absorption coefficient is given by
\begin{equation}\label{alpha}
\alpha'_{\nu}\approx{3 \gamma_c n' (P'_{e,syn}/\nu'_{syn})\over 16\pi m_e\nu^{5/3}\nu_{b0}^{1/3}}
\left(\frac{16 m_e c\nu}{3 q B_b}\right)^{-4/3}\ .
\end{equation}
We solve $\alpha'_{\nu'_{sa}}\Delta'=1$ for $\nu'_{sa}$, where $\Delta'\approx
R/\Gamma$ is the width of the shell in the local frame, and then we have
$\nu_{sa}=\Gamma\nu'_{sa}$. The synchrotron frequencies are given by
\begin{eqnarray}
\label{nu_0}
\nu_{c}&= & 3.7\times 10^{10}\,(1+Y)^{-2}\epsilon_e^{3/2}
\epsilon_B^{-3/2}L_{52}^{-3/2}\Gamma_{2.5}^8 t_{v,-2}\;{\rm Hz}\ ,
\nonumber \\ \label{nu_c}
\nu_{0}&= & 7.8\times 10^{13}\,\epsilon_e^{-1/2}
\epsilon_B^{1/2}L_{52}^{1/2}\Gamma_{2.5}^{-2}t_{v,-2}^{-1}\;{\rm Hz}\ ,
\nonumber \\ \label{nu_ac}
\nu_{ac}&= & 5.7\times 10^{13}f_p^{-8/5}(1+Y)^{-3/5}\epsilon_e^{-9/5}
\epsilon_B^{-2/5}L_{52}^{1/5}\Gamma_{2.5}^{-1/5}t_{v,-2}^{-2/5}\;{\rm Hz}\ ,
\nonumber \\ \label{nu_sa}
\nu_{sa}&= & 2.0\times 10^{16}\,(1+Y)^{-1/3}\epsilon_e^{-1/3}L_{52}^{1/3}
\Gamma_{2.5}^{-1}t_{v,-2}^{-2/3}\;{\rm Hz}\ ,
\\ \label{nu_m}
\nu_{m}&= & 2.9\times 10^{19}\,f_p^2(1+Y)^{-1/3}\epsilon_e^{3/2}
\epsilon_B^{1/2}L_{52}^{1/2}\Gamma_{2.5}^{-2}t_{v,-2}^{-1}\;{\rm Hz}\ ,
\nonumber \\ \label{nu_M}
\nu_{M}&= & 2.3\times 10^{24}\,(1+Y)^{-1}\Gamma_{2.5}\;{\rm Hz}\ ,
\nonumber
\end{eqnarray}
and the synchrotron spectrum is\footnote{If there is significant
mixing of the shocked fluid (which may be the case if there is strong turbulence)
then we will have $F_\nu\propto\nu^2$ immediately below $\nu_{sa}$, instead of
$F_\nu\propto\nu^{11/8}$ for $\nu_{ac}<\nu<\nu_{sa}$ and $F_\nu\propto\nu^{2}$
below $\nu_{ac}=\nu_{sa}(\nu_c/\nu_m)^{4/5}$ (Granot, Piran, \& Sari 2000).
However, this will not have any significant
effect on the SSC spectrum, and since $\nu_{sa}$ is typically below the
observational window, it will be quite hard to distinguish between these two
possibilities observationally.}
\begin{equation}\label{Fnu_syn_gc<1}
{\nu F_{\nu}\over\nu_{m}F_{\nu_m}}=\left\{\matrix{
(\nu_{sa}/\nu_{m})^{1/2}(\nu_{ac}/\nu_{sa})^{19/8}
(\nu/\nu_{ac})^3 & \ \ \nu<\nu_{ac} \cr & \cr
(\nu_{sa}/\nu_{m})^{1/2}(\nu/\nu_{sa})^{19/8}
& \ \ \nu_{ac}<\nu<\nu_{sa} \cr & \cr
(\nu/\nu_{m})^{1/2} & \nu_{sa}<\nu<\nu_{m} \cr & \cr
(\nu/\nu_{m})^{(2-p)/2} & \nu_{m}<\nu<\nu_{M}}\right. \ ,
\end{equation}
where from the normalization condition we obtain,
\begin{eqnarray}\label{norm}
{L_{iso}\over 4\pi D^2}&\hspace{-0.3cm} =\hspace{-0.3cm}&\int_0^\infty d\nu F_{\nu}
=6f_p^{-1}\nu_m F_{\nu_m}\ ,
\\
\nu_m F_{\nu_m}&\hspace{-0.3cm}=\hspace{-0.3cm}&{f_p L_{iso}\over 24\pi D^2}=
1.3\times 10^{-6}f_p L_{52}D_{28}^{-2}\;{\rm erg\,\,cm^{-2}\,s^{-1}}\ ,
\end{eqnarray}
where $D=10^{28}D_{28}\;{\rm cm}$ is the distance to the GRB.

The SSC spectrum is given by
\begin{equation}\label{Fnu_SSC}
{\nu F_{\nu}^{SC}\over Y\nu_{m}F_{\nu_m}}=\left\{\matrix{
(\nu_{sa}^{SC}/\nu_{m}^{SC})^{1/2}(\nu/\nu_{sa}^{SC})^{2}
& \ \ \nu<\nu_{sa}^{SC} \cr & \cr
(\nu/\nu_{m}^{SC})^{1/2} & \nu_{sa}^{SC}<\nu<\nu_{m}^{SC} \cr & \cr
(\nu/\nu_{m}^{SC})^{(2-p)/2} & \nu_{m}^{SC}<\nu<\nu_{KN}^{SC}\cr & \cr
(\nu_{KN}^{SC}/\nu_{m}^{SC})^{2-p\over 2}(\nu/\nu_{KN}^{SC})^{1-2p\over 2}
 & \nu_{KN}^{SC}<\nu<\nu_{M}^{SC}\cr & \cr
}\right. \ ,
\end{equation}
where $\nu_{sa}^{SC}\equiv\max(\gamma_c^2,1)\nu_{sa}$ and
\begin{eqnarray}
\nu_{m}^{SC}&\hspace{-0.35cm}= \hspace{-0.3cm}&\gamma_m^2\nu_m= 1.1\times 10^{25}f_p^4
\epsilon_e^{7/2}\epsilon_B^{1/2}L_{52}^{1/2}\Gamma_{2.5}^{-2}t_{v,-2}^{-1}\;{\rm Hz}\ ,
\nonumber \\
\nu_{KN}^{SC}&\hspace{-0.35cm}= \hspace{-0.3cm}&\Gamma^2 m_e^2 c^4/h^2\nu_m = 5.2\times 10^{25}
f_p^{-2}\epsilon_e^{-3/2}\epsilon_B^{-1/2}L_{52}^{-1/2}\Gamma_{2.5}^{4}t_{v,-2}\;{\rm Hz}\ ,
\label{freq} \\
\nu_{M}^{SC}&\hspace{-0.35cm}= \hspace{-0.3cm}&\Gamma\gamma_{M}m_e c^2/h = 6.6\times 10^{27}
(1+Y)^{-1/2}\epsilon_e^{1/4}\epsilon_B^{-1/4}L_{52}^{-1/4}
\Gamma_{2.5}^{5/2}t_{v,-2}^{1/2}\;{\rm Hz}\ .
\nonumber
\end{eqnarray}
If $\nu_{KN}^{SC}<\nu_{m}^{SC}$ then we have $\nu F_\nu\propto\nu^{1/2-p}$
for $\nu_{KN}^{SC}<\nu<\nu_{M}^{SC}$. For details about the spectrum above the Klein-Nishina
frequency, $\nu_{KN}^{SC}$, see Guetta \& Granot (2002).

%%%%%%%%%%%%%%%%%%%%%%%%%%%%%%%%%%%%%%%%%%%%%%%%%%%%%%%%%%%%%%%%
\section{The High Energy Cutoff}
\label{limits}

In order for high energy photons to escape the system and reach the observer,
they must overcome a few potential obstacles along the way.
Inside the source, there are two main constraints: i) the opacity of the high energy photons to
pair production due to interaction with lower energy photon, $\tau_{\gamma\gamma}$, must be
smaller than 1 in order for high energy photons to escape, and
ii) the Thomson optical depth due to pair production, $\tau_p$, must be smaller than
unity in order for the observed prompt GRB emission to escape and reach the observer.
These effects have been studied by several different authors (Sari \& Piran 1997;
Lithwick \& Sari 2001; Guetta, Spada \& Waxman 2001). Outside of the source, high energy
photons ($\gtrsim 500\;$GeV) may interact with photons from the cosmic IR background to
produce pairs (Salamon \& Stecker 1998; Dai \& Lu 2002).

We reparameterize the expressions of Lithwick \& Sari (2001) for $\tau_{\gamma\gamma}$ and $\tau_p$,
using our parameters, and express the requirements $\tau_{\gamma\gamma}<1$ and $\tau_p<1$ as
constraints on the Lorentz factor $\Gamma$:
\begin{eqnarray}
\label{Gamma_gg}
\Gamma&\hspace{-0.25cm}> \hspace{-0.25cm}&69\,\epsilon_e^{3(p-2)\over8(p+1)}\epsilon_B^{(p-2)\over 8(p+1)}
L_{52}^{(p+2)\over 8(p+1)}t_{v,-2}^{-{p\over 4(p+1)}}\varepsilon_{\rm max}^{{p\over 4(p+1)}}
\quad (\tau_{\gamma\gamma}<1)\ ,
\\ \label{Gamma_p}
\Gamma&\hspace{-0.25cm}> \hspace{-0.25cm}&170\,\epsilon_e^{3(p-2)\over 2(3p+4)}\epsilon_B^{(p-2)
\over 2(3p+4)}  L_{52}^{(p+2)\over 2(3p+4)}t_{v,-2}^{(p+2)\over (3p+4)}  \quad\quad\ \  (\tau_p<1)\ ,
\end{eqnarray}
where $\varepsilon_{\rm max}$ is the maximal energy of a photon that can escape (measured in the
observer frame, in units of $m_ec^2$) and the numerical coefficients are for $p=2.5$, but do not
vary by more than $10\%$ for $2.05<p<2.9$. In order to see the soft gamma-rays from the prompt GRB,
the inequality in Eq. (\ref{Gamma_p}) must be satisfied, in which case there will be an upper
cutoff at
\begin{equation}\label{e_max}
h\nu_{\gamma\gamma}\equiv\varepsilon_{\rm max}m_ec^2=2.6\,\epsilon_e^{-3(p-2)\over 2p}
\epsilon_B^{(2-p)\over 2p}L_{52}^{-{(p+2)\over 2p}}\Gamma_{2.5}^{4(p+1)\over p}t_{v,-2}\;{\rm GeV}\ ,
\end{equation}
where the numerical coefficient is for $p=2.5$ and varies by less than $10\%$ for $2.17<p<2.67$.
The maximum photon energy detected by EGRET during the prompt emission is $\sim 3$ GeV
(Hurley et al. 1994), for which Eq. (\ref{Gamma_gg}) implies $\Gamma>370$ for
our fiducial parameters.

In Figure \ref{fig1} we show the synchrotron and SSC $\nu F_\nu$ spectra, for different
values of $\Gamma$ and $t_v$. As can be seen from this figure, the high energy cutoff is
typically determined by the opacity to pair production, and is given by Eq. (\ref{e_max}).
As mentioned above, the high energy photons ($\gtrsim 500\;{\rm GeV}$) that escape the
source may still interact with the cosmic IR background and produce pairs.
However, as illustrated in Figure \ref{fig1}, in order for photons in this energy range to
leave the system, one needs $\Gamma\gtrsim 600$ and $t_v\gtrsim 0.01\;{\rm s}$, which,
in turn, implies $h\nu_m=E_p\lesssim 1\;{\rm keV}$. Therefore, this effect is expected to
be irrelevant for typical GRBs, with $E_p$ in the BATSE range, and might play a role only for
the low end of the $E_p$ distribution of the X-ray flashes.

It has recently been claimed (Dai \& Lu 2002) that a delayed
emission on a timescale of $\Delta t\sim 10^3\;{\rm sec}$ after
the GRB may result due to the inverse Compton upscattering of the
CMB by the pairs produced by photons with energies $\gtrsim
300\;{\rm GeV}$ that are emitted during the prompt GRB, with the
cosmic IR background. The pairs are produced at a typical distance
of $R_{pair}\approx 5.8\times 10^{24}\;{\rm cm}$, and loose their
energy by upscattering CMB photons over a length scale of
$R_{IC}\approx 7.3\times 10^{24}\;{\rm cm}$. Dai \& Lu estimated
the delay time by $\Delta t\sim R/2\gamma_e^2$, where
$R=\max(R_{pair},R_{IC})$. This estimate is based on the fact that
the pairs that are produced initially propagate almost in the same
direction as the high energy photon (i.e. in the radial direction
from the GRB) and on the assumption that they do not change their
direction by more than $1/\gamma_e$ over $R_{IC}$. However, the
presence of intergalactic magnetic fields (IGMF), $B_{IG}$, at
$R_{pair}$ will cause a deflection angle of
\begin{equation}\label{theta_def}
\theta_{\rm def}={R_{IC}\over R_L}=1.4\times 10^5\left({B_{IG}\over 10^{-11}\,{\rm G}}\right)
\left({\gamma_e\over 3\times 10^5}\right)^{-2}\ ,
\end{equation}
where $R_L=\gamma_e m_e c^2/qB$ is the Larmor radius of the
electron. In order for the above estimate for $\Delta t$ not to be
effected by this deflection, we need $\theta_{\rm
def}<1/\gamma_e$, which according to Eq. (\ref{theta_def}) implies
$B_{IG}<B_0=2.3\times 10^{-22}\;{\rm G}$. For $B_{IG}>B_0$, the
value of $\Delta t$ increases by a factor of $\sim(B_{IG}/B_0)^2$
compared to the estimate of Dai \& Lu. For $\theta_{\rm def}>1$,
i.e. $B_{IG}\gtrsim 10^{-16}\;{\rm G}$, we expect $\Delta t\sim
R_{IC}/c\sim 10^7\;{\rm yr}$, and a roughly isotropic emission.
Therefore, the detection of such a delayed emission will suggest
an IGMF $\lesssim 10^{-21}\;$G at a distance of a few Mpc from the
site of the GRB, while the lack of detection of this emission will
imply a larger IGMF. As in such a distance from the site of the
GRB one typically expects to reach a void, this can serve as a
mothod for estimating the highly uncertain value of the IGMF in
voids. A similar suggestion was made by Plaga (1995).

%%%%%%%%%%%%%%%%%%%%%%%%%%%%%%%%%%%%%%%%%%%%%%%%%%%%%%%%%%%%%%%%
\section{Discussion}
\label{sec:dis}

We have calculated the synchrotron and SSC emission during the prompt GRB
from internal shocks, and studied the relation between these two components
and its dependence on the model parameters. Our analysis takes into account
the high energy cutoffs due to the Klein-Nishina effect, pair production
with low energy photons and with the cosmic IR background, and Thomson
scattering.

For $p>2$ the emission above $\sim 100\;{\rm MeV}$, is typically dominated by the SSC emission,
while the synchrotron component is dominant at lower energies. If the variability time is
$t_v\gtrsim 1\;{\rm ms}$, then $E_p\sim 300\;{\rm keV}$ would imply a cutoff at
$\sim 30\;{\rm MeV}$, and the synchrotron emission would be dominant at all energies.
Future observations by GLAST may help determine the value of
$p$ and whether the high energy emission is consistent with a single power law
or has a break where the $\nu F_\nu$ slope turns from negative to positive.
The former would imply that the high energy emission is dominated by a single component--
the synchrotron emission, while the latter implies that the SSC
component becomes dominant above a certain energy ($\sim 100\;$MeV).

The SSC high energy emission should show a similar variability to
that observed in the BATSE range. An additional emission mechanism
that might contribute to the high energy emission, is external
Compton, which may be relevant if GRBs occur inside pulsar wind
bubbles (Guetta \& Granot 2002).

In addition, there might be delayed high energy emission from the
internal shocks, due to upscattering, aas was suggested by  of the
CMB by $e^\pm$ pairs produced by the interaction between $\gtrsim
300\;$GeV photons from the prompt GRB with the IR background
photons (Dai \& Lu 2002). The detection of such emission would be
possible only for inter galactic magnetic fields (IGMF) $\lesssim
10^{-21}\;$G at a distance of a few Mpc from the site of the GRB,
so that it may serve a probe for the strength of the IGMF in
voids.

\begin{figure}
\centering
\noindent
\includegraphics[width=13cm]{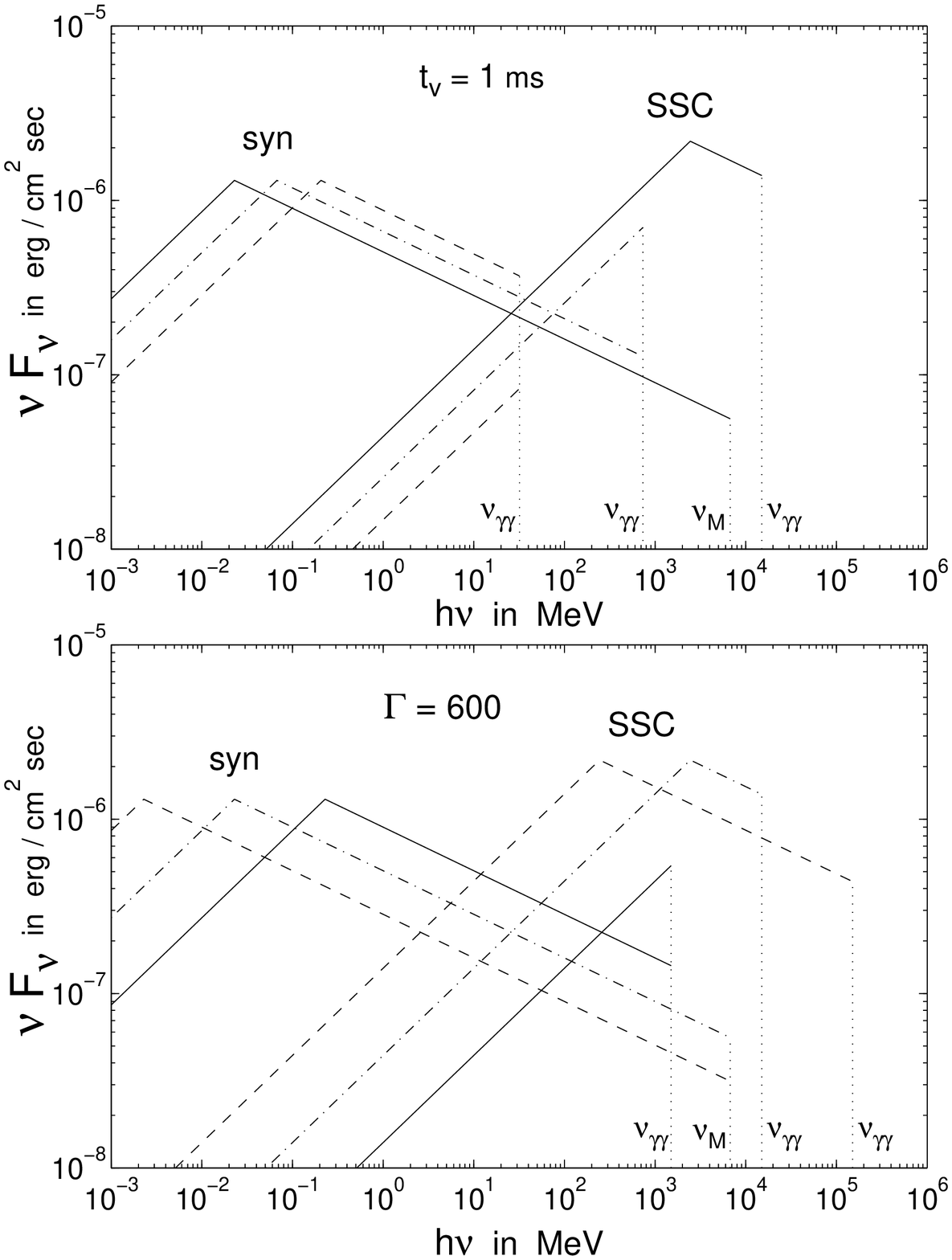}
%\plotone{f1.eps}
\caption{\label{fig1} The $\nu F_\nu$ spectrum from the prompt GRB
for the synchrotron and SSC components, including the relevant
high energy cutoff among the possibilities discussed in the text.
In the upper panel we fix $t_v=1\;{\rm ms}$, and the dashed,
dashed-dotted and solid lines correspond to $\Gamma=200,\,350$ and
$600$, respectively. In the lower panel we fix $\Gamma=600$, and
the  dashed, dashed-dotted and solid lines correspond to
$t_v=10,\,1$ and $0.1\;{\rm ms}$, respectively. In both panels we
use $L_{52}=1$ and $p=2.5$, as well as $\epsilon_e=0.45$ and
$\epsilon_B=0.1$, where the latter were chosen so that the peak of
$\nu F_\nu$ of the synchrotron would be at the right range
(Guetta, Spada \& Waxman 2001). }
\end{figure}

As can be seen from Eqs. (\ref{nu_c}), (\ref{e_max}) and Figure \ref{fig1},
larger values of $\Gamma$ or $t_v$ shift the cutoff at $h\nu_{\gamma\gamma}$
to larger energies, while at the same time, it implies lower values of $E_p$.
For example, in order to have $\sim 1\;{\rm GeV}$ photons for
$t_v\sim 1\;{\rm ms}$ we need $\Gamma\gtrsim 350$, which in turn, imply
$E_p\lesssim 100\;{\rm keV}$. If X-ray flashes are GRBs with such parameters,
as suggested by Guetta, Spada \& Waxman (2001), then we can expect GeV emission
from X-ray flashes.

In order to explain the prompt high energy photons,
of up to $\sim 3\;{\rm GeV}$, that were observed in GRB 940217 (Hurley et al. 1994),
together with the value of $E_p\approx 200\;{\rm keV}$ that was measured for this burst,
we need a very small variability time $t_v\sim 0.1\;{\rm ms}$, and $\Gamma\sim 600$
(see the solid curve in the lower panel of Figure \ref{fig1}). If indeed such high energy
emission is typical for GRBs with $E_p$ in the BATSE range, as can be tested by the future
mission GLAST, this might suggest very low variability times ($t_v\lesssim 0.1\;{\rm ms}$).
This possibility is consistent with the fact that in many GRBs the shortest measured
variability time is limited by the temporal resolution of the instrument, and there is no
observational lower limit on $t_v$. On the other hand, $t_v\sim 0.1\;{\rm ms}$ implies a
source size $\lesssim ct_v\sim 30\,t_{v,-4}\;{\rm Km}$, so that it is unlikely that $t_v$
can be much smaller that $0.1\;{\rm ms}$. Therefore, this might imply a typical variability
time of $t_v\sim 0.1\;{\rm ms}$, which may be resolved by super AGILE.

\acknowledgements We thank Eli Waxman for his useful comments.
This research was supported by the partial support of the Italian
Ministry for University and Research (MIUR) through the grant
Cofin-01-02-43 (DG) and by the grant NSF PHY 00-70928 (JG). DG
thanks the Institute for Advanced Study, where this research was
carried out, for the hospitality and the nice working atmosphere.

\newpage

\end{document}